\documentclass[12pt]{article}
\usepackage[english]{babel}
\usepackage{graphicx}
\usepackage{mathrsfs}
\usepackage{amsfonts}
\usepackage{fancyhdr}
\usepackage{textcomp}
\usepackage{psfrag}
\usepackage[official]{eurosym}
\newtheorem{theorem}{Theorem}

\newcommand\cC{{\cal C}}

\newcommand\cF{{\cal F}}

\newcommand\cT{{\cal T}}
\newcommand\cU{{\cal U}}

\newcommand\rd{\rm d}
\newcommand\p{\partial}

\newcommand\bX{{\bf X}}
\newcommand\bt{{\hat{\bf t}}}
\newcommand\bu{{\bf u}}
\newcommand\bx{{\bf x}}
\newcommand\bom{{\mbox{\boldmath $\omega$}}}

\newcommand\ba{{\bf a}}

\newcommand\bb{{\hat{\bf b}}}

\newcommand\NN{{\hbox{I\kern-.14em{N}}}}
\newcommand\RR{{\hbox{I\kern-.14em{R}}}}
\newcommand\ZZ{{\hbox{I\kern-.14em{Z}}}}

\begin{document}
\begin{center}
{\Large 
{\bf Velocity, energy and helicity of vortex knots and unknots }
}

\vskip 1.0cm

{ F. Maggioni\footnotemark[1]},
{ S. Alamri\footnotemark[2]},
{ C.F. Barenghi\footnotemark[3]}
and
{R.L. Ricca\footnotemark[4]}.
\end{center}


\footnotetext[1]{\small
\noindent Dept. Mathematics, Statistics, Computer Science and 
Applications, U. Bergamo\\
Via dei Caniana 2, 24127 Bergamo, ITALY}

\footnotetext[2]{\small
\noindent Dept. Applied Mathematics, College of Applied 
Science, U. Taibah\\
P.O. Box 344, Al-Madinah Al-Munawarah, SAUDI ARABIA}  

\footnotetext[3]{\small \noindent School of Mathematics and Statistics, U. Newcastle\\ 
Newcastle upon Tyne, NE1 7RU, U.K.}

\footnotetext[4]{\small \noindent Dept. Mathematics and Applications, U. Milano-Bicocca\\ 
Via Cozzi 53, 20125 Milano, ITALY}


\begin {abstract}
\noindent In this paper we determine the velocity, the energy and estimate 
writhe and twist helicity contributions of vortex filaments in the 
shape of torus knots and unknots (toroidal and poloidal coils) in 
a perfect fluid. Calculations are performed by numerical integration 
of the Biot-Savart law. Vortex complexity is parametrized by the 
winding number $w$, given by the ratio of the number of meridian 
wraps to that of the longitudinal wraps. We find that for $w<1$ 
vortex knots and toroidal coils move faster and carry more energy 
than a reference vortex ring of same size and circulation, whereas 
for $w>1$ knots and poloidal coils have approximately same speed 
and energy of the reference vortex ring. Helicity is dominated by 
the writhe contribution. Finally, we confirm the stabilizing effect 
of the Biot-Savart law for all knots and unknots tested, that are 
found to be structurally stable over a distance of several diameters. 
Our results also apply to quantized vortices in superfluid $^4$He.
\end{abstract}




\section{Introduction}
The study of vortex filament motion in an ideal fluid, and in 
particular of vortex rings in presence or absence of periodic 
displacements of the vortex axis from the circular shape (Kelvin 
waves), dates back to the late 1800s \cite{ref:Ke80,ref:JJ83}. 
Despite its long history, this study is still an active area of 
research \cite{ref:Fu09,ref:Ha09,ref:FM09}. For example, 
the fact that upon large amplitude Kelvin waves vortex rings can 
slow down and even reverse their translational motion has been 
recognised only recently \cite{ref:KiMa02,ref:Ba06}. 
Alongside the traditional interest for problems in classical 
fluid mechanics, additional interest is motivated by current 
work on superfluid helium  
\cite{ref:Wea07,ref:BS09,ref:Go09,ref:Sreeni09} and atomic 
Bose-Einstein condensates \cite{ref:TK08,ref:Be09,ref:Hu09}. 

Here we shall be concerned with slightly more complex vortex 
structures; namely vortex filaments in the shape of torus knots 
and unknots in an ideal fluid. Since these vortices are closely 
related to circular vortex filaments with small-amplitude 
distortions lying on a mathematical torus, like vortex rings 
they propagate in the fluid by self-induction along the central 
axis of the torus, and they also rotate in their meridian plane 
(the poloidal plane of the torus) as their vortex core spins 
around the local center of mass, inducing an additional twisting 
motion of the vortex on itself, that cannot be neglected. 

Among all possible knot types, torus knots constitute a special 
family of knots amenable to particularly simple mathematical 
description. These knots can be described by closed curves wound
on a mathematical torus a number of times in the longitudinal 
direction of the torus and a number of times in the meridian 
direction. If these two numbers are co-prime integers greater than 
one, then we have standard torus knots (see Figure~\ref{knots}), 
whereas if the curve winds the torus only once in one of the two 
directions, then the curve is simply unknotted (see 
Figure~\ref{unknots}), reducing to the 
standard circle when both numbers coincide with one. Even though 
torus unknots have trivial topology, their geometry may be 
rather complex, taking the shape of toroidal or poloidal coil,
depending along which direction the curve is multiply wound. 

Vortex torus knots and unknots provide a good example of structures 
that are relatively complex in space, but still amenable to study
relationships between dynamical properties, like velocity and energy,
and geometric and topological features, a step further towards the 
study of more complex structures present in turbulent flows.
The aim of this paper is thus to continue and extend previous work
\cite{ref:RSB99,ref:MABR09} to more complex vortex structures.
Here we shall investigate the propagation velocity and the kinetic 
energy of vortex torus knots and unknots in some generality, by comparing 
results to a standard vortex ring of same size. Since superfluid $^4$He 
has zero viscosity, thus providing a realistic example of an Euler fluid, 
we shall choose circulation and vortex core radius as physically realistic 
quantities for quantized vortices in superfluid helium, and we shall carry 
out the research by direct numerical integration of the Biot-Savart law.
Unlike previous vortex dynamics calculations of quantized vorticity
\cite{ref:Sch88,ref:Aa94,ref:BBSD97,ref:TBA04}, however, we shall assume 
that no friction force \cite{ref:BDV82} acts on the superfluid vortices; 
thus our results will apply to superfluid helium at temperatures below 
$1~\rm K$, where the dissipative effects of the normal fluid are truly 
negligible.

\section{Mathematical background}
\subsection{Vortex motion under Biot-Savart and LIA law}
We consider vortex motion in an ideal, incompressible fluid, 
in an unbounded domain. The velocity field $\bu=\bu(\bx,t)$, smooth 
function of the vector position $\bx$ and time $t$, satisfies
\begin{equation}
    \nabla\cdot\bu=0\quad \textrm{in}\ \RR^3\ ,\qquad \bu\to 0\quad 
    \textrm{as}\ \bx\rightarrow\infty\ ,
    \label{bu}
\end{equation}
with vorticity $\bom$ defined by
\begin{equation}
    \bom=\nabla\times\bu\ ,\qquad 
    \nabla\cdot\bom=0\quad \textrm{in}\ \RR^3\ .
    \label{bom}
\end{equation}
In absence of viscosity, fluid evolution is governed by the Euler
equations and vortex motion obeys Helmholtz's conservation laws
\cite{ref:Sa92}. Transport of vorticity is given by 
\begin{equation}
    \frac{\partial\bom}{\partial t}
    =\nabla\times\left(\bu\times\bom\right)\ ,
    \label{vortrans}
\end{equation}
admitting formal solutions in terms of the Cauchy equations
\begin{equation}
    \omega_i(\bx,t)=
    \omega_j(\ba,0)\frac{\partial x_i}{\partial{\rm a}_j}\ .
    \label{cauchy}
\end{equation}
From this expression we can see how both convection of vorticity 
from the initial position $\ba$ to the final position $\bx$, and 
the simultaneous rotation and distortion of the vortex elements 
by the deformation tensor $\p x_i/\p{\rm a}_j$ are combined together.  
Since this tensor is associated with a continuous deformation of the
vortex elements (by the diffeomorphism of the flow map), vorticity 
is thus mapped continuously from its initial configuration $\bom(\ba,0)$ 
to the final state given by $\bom(\bx,t)$; hence the Cauchy equations 
establish a topological equivalence between initial and final 
configuration by preserving vorticity topology. In 
absence of dissipation, physical properties such as kinetic energy, 
helicity and momenta are therefore conserved along with topological 
quantities such as knot type, minimum crossing number and self-linking 
number \cite{ref:RB96}.

The kinetic energy per unit density $E$ is given by  
\begin{equation}
    E=\frac{1}{2}\int_{V}\|\bu\|^2\,\rd^{3}\bx=constant\ ,
    \label{kinetic}
\end{equation}
where $V=V$ is the fluid volume, and the kinetic helicity $H$ by
\begin{equation}
    H=\int_{V}\bu \cdot \bom\,\rd^{3}\bx=constant\ .
    \label{helicity}
\end{equation}
 Here we assume to have only one vortex 
filament $\cF$ in isolation, where $\cF$ is centred on the curve $\cC$ of 
equation $\bX=\bX(s)$ ($s$ being the arc-length of $\cC$).  
The filament axis $\cC$ is given by a smooth (that is at least $C^{2}$),
simple (i.e. without self-intersections), space curve $\cC$.  The
filament volume is given by $V(\cF)=\pi a^{2}L$, where $L=L(\cC)$ is
the total length of $\cC$ and $a$ is the radius of the vortex core, 
assumed to be uniformly circular all along $\cC$ and much smaller than any
length scale of interest in the flow (thin-filament approximation); this
assumption is relevant (and particularly realistic) in the context of 
superfluid helium vortex dynamics, where typically $a \approx 10^{-8}\rm cm$.

Vortex motion is governed by the Biot-Savart law (BS for short) given by
\begin{equation}
  \bu(\bx)=\frac{\Gamma}{4\pi}\oint_{\cC}
  \frac{\bt\times(\bx-\bX(s))}{\|\bx-\bX(s)\|^{3}}\,\rd s\ ,
  \label{bs}
\end{equation}
where $\Gamma$ is the vortex circulation due to $\bom=\omega_{0}\bt$,
where $\omega_{0}$ is a constant and $\bt=\bt(s)=\rd\bX/\rd s$ 
the unit tangent to $\cC$.  Since the
Biot-Savart integral is a global functional of vorticity and geometry, 
analytical solutions in closed form other than the classical solutions 
associated with rectilinear, circular and helical geometry are very 
difficult to obtain. Considerable analytical progress, however, has 
been done by using the Localized Induction Approximation (LIA for short) 
law.  This equation, first derived by Da Rios \cite{ref:DR06} and 
independently re-discovered by Arms \& Hama \cite{ref:AH65} (see the 
review by Ricca \cite{ref:Ri96}), is obtained by a Taylor's expansion 
of the Biot-Savart integrand from a point on $\cC$ (see, for instance, 
the derivation by Batchelor \cite{ref:Ba67}). By neglecting the rotational 
component of the self-induced velocity (that in any case does not contribute 
to the displacement of the vortex filament in the fluid) and non-local terms, 
the LIA equation takes the simplified form 
\begin{equation}
  \bu_{\rm LIA}=\frac{\Gamma c}{4\pi}\ln{\delta}\,\bb\propto c\bb\ ,
  \label{lia}
\end{equation}
where $c=c(s)$ is the local curvature of $\cC$, $\delta$ is a measure 
of the aspect ratio of the vortex, given by the radius of curvature 
divided by the vortex core radius) and $\bb=\bb(s)$ the unit binormal 
vector to $\cC$. 

\begin{figure}[t!]
\centering
\includegraphics[width=0.7\textwidth]{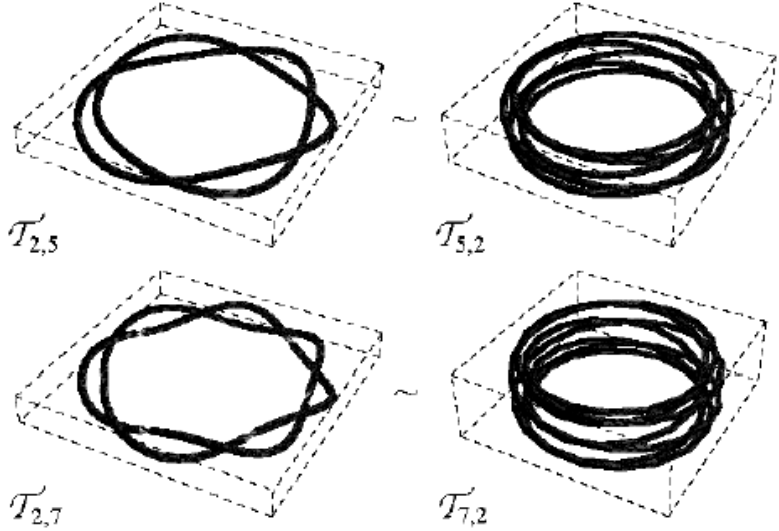}
\caption{Examples of torus knots with winding number $w>1$ (left column)
and $w<1$ (right column). Note that for given $p$ and $q$, torus knot 
$\cT_{p,q}$ is topologically equivalent to the knot $\cT_{q,p}$, and we 
write $\cT_{p,q}\sim\cT_{q,p}$. The tube centred on each knot is only 
for visualization purposes.}
\label{knots}
\end{figure}

\subsection{Torus knots}
\begin{figure}[t!]
\centering
\includegraphics[width=0.7\textwidth]{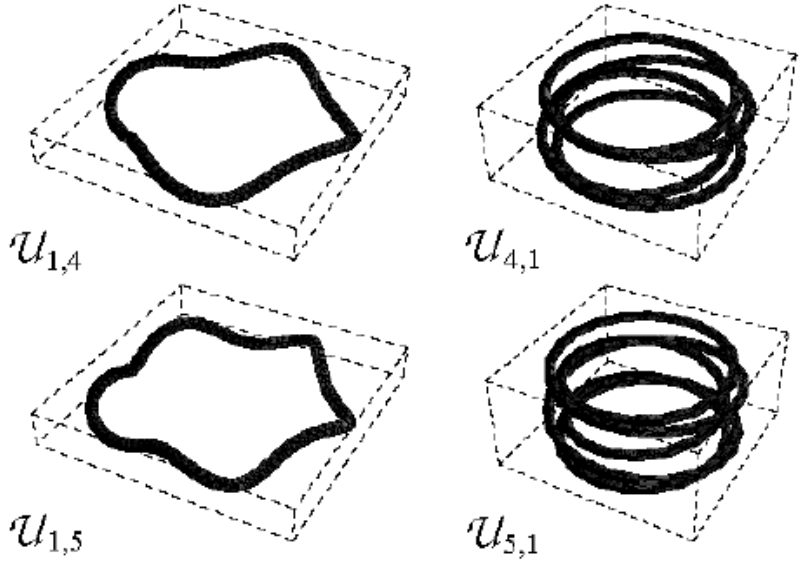}
\caption{Examples of torus unknots with winding number $w>1$ (poloidal 
coils; left column) and $w<1$ (toroidal coils; right column). All these 
unknots are topologically equivalent to the standard circle. The tube 
centred on each unknot is shown only for visualization purposes.}
\label{unknots}
\end{figure}

We consider a particular family of vortex configurations in the shape
of torus knots in $\RR^{3}$.  These are given when the curve $\cC$ 
takes the shape of a torus knot $\cT_{p,q}$ ($\{p,q\}$ co-prime integers
with $p>1$ and $q>1$), given by a closed curve wound on a mathematical 
torus $p$ times in the longitudinal (toroidal) direction and $q$ times 
in the meridian
(poloidal) direction (see Figure~\ref{knots}).  When one of the integers 
is equal to one and the other is equal to $m$ the curve is no longer 
knotted, thus forming an unknot homeomorphic to the standard circle 
$\cU_0$, but with a more complex geometry. Depending on which index 
takes the value $m$, the curve takes the shape of a toroidal
coil ($p=m$) $\cU_{m,1}$, or a poloidal coil ($q=1$)
$\cU_{1,m}$ (see Figure~\ref{unknots}).  When $\{p,q\}$ are both rational, 
$\cT_{p,q}$ is no longer a closed knot, the curve covering the toroidal 
surface completely. Here we shall consider only curves given by $\{p,q\}$ 
integers.  The ratio $w=q/p$
denotes the \emph{winding number} and $Lk=pq$ the \emph{self-linking
number}, two topological invariants of $\cT_{p,q}$.  Note that for
given $p$ and $q$ the knot $\cT_{p,q}$ is topologically equivalent to
$\cT_{q,p}$, that is $\cT_{p,q}\sim\cT_{q,p}$, i.e. they are the same
knot type, even though their geometry is completely different: toroidal
coils resemble thick vortex rings with one full turn of twist of vorticity 
(internal swirl flow), whereas poloidal coils may recall circular jet flow
generated by helicoidal distribution of vorticity on circular central axis. 

A useful measure of geometric complexity of $\cC$ is given by the 
\emph{writhing number} \cite{ref:Fu71} defined by
\begin{equation}
    Wr\left(\cC\right)\equiv\frac{1}{4\pi}\oint_{\cC}\oint_{\cC}
    \frac{\bt(s)\times\bt(s^{\ast})\cdot 
    [\bX(s)-\bX(s^{\ast})]}{\|\bX(s)-\bX(s^{\ast})\|^3}
   \,\rd s \ \rd s^{\ast}\ ,
    \label{writhe}
\end{equation}
\textnormal{where $\bX(s)$ and $\bX(s^\ast)$ denote two points 
on the curve $\cC$ for any pair $\{s,s^\ast\}\in \left[0,L\right]$, 
the integration being performed twice on the same curve $\cC$.} The 
writhing number provides a direct measure of coiling and distortion 
of the filament in space. This information can be related to the 
total twist $Tw$ and, by knowing the self-linking number $Lk$ of each 
knot, to the helicity $H$ of the vortex, that can be estimated to be
$H=\Gamma Lk=\Gamma(Wr+Tw)$ \cite{ref:RM92}, without resorting to direct 
calculation of the integral (\ref{helicity}).

Now, let us identify each knot $\cT_{p,q}$ with the vortex filament
(we shall hereafter refer to $\cT_{p,q}$ as the vortex torus knot), 
and consider dynamics and energy by using the BS (\ref{bs}) and the 
LIA law (\ref{lia}) to determine the effects of different 
geometries and topologies on dynamical properties and kinetic energy.

The existence of torus knot solutions to LIA were found by Kida 
\cite{ref:Ki81} in terms of elliptic integrals. By re-writing LIA 
in cylindrical polar coordinates $(r,\alpha, z)$, and by using 
standard linear perturbation techniques, small-amplitude torus knot 
solutions (asymptotically equivalent to Kida's solutions) were derived 
by Ricca \cite{ref:Ri93}. These latter give solution curves explicitly 
in terms of the arc-length $s$, given by
\begin{equation}
\left\{ 
\begin{array}{lcl}
  r= r_0 +\epsilon\sin(w\phi)\ ,\\[2mm]
  \alpha= \displaystyle{\frac{s}{r_0} +\frac{\epsilon}{wr_0}
    \cos(w\phi)}\ , \\[2mm]
  z=\displaystyle{\frac{t}{r_0}
  +\epsilon\left(1+\frac{1}{w^2}\right)^{1/2}
    \cos(w\phi)}\ ,
\end{array} 
\right.
\label{sols}
\end{equation}
where $r_0$ is the radius of the torus circular axis and
$\epsilon\ll1$ is the inverse of the aspect ratio of the vortex.
Since the LIA is related \cite{ref:Ri96} to the one-dimensional
Non-Linear Schr\"{o}dinger Equation (NLSE), torus knot solutions 
(\ref{sols}) correspond to helical travelling waves propagating 
along the filament axis, with wave speed $\kappa$ and phase 
$\phi=(s-\kappa t)/r_0$. Vortex motion is given by a rigid body 
translation and rotation, with translation velocity $u=\dot{z}=u$ 
along the torus central axis and a uniform helical motion along 
the circular axis of the torus given by radial and
angular velocity components $\dot{r}$ and $\dot{\alpha}$.
In physical terms, these waves provide an efficient mechanism 
for the transport of kinetic energy and momenta throughout the fluid.

By using eqs. (\ref{sols}), Ricca \cite{ref:Ri95} proved the 
following linear stability result:

\begin{theorem}  
Let $\cT_{p,q}$ be a small-amplitude vortex torus knot under LIA. 
Then $\cT_{p,q}$ is steady and stable under linear perturbations 
if and only if $q>p$ ($w>1$).
\label{Ricca2005}
\end{theorem}

This result provides a criterium for LIA stability of vortex knots,
and it can be easily extended to inspect stability of torus unknots 
(i.e. toroidal and poloidal coils). This stability result has been 
confirmed for the knot types tested by numerical experiments 
\cite{ref:RSB99}. Interestingly, LIA unstable torus knot were 
found to be stable under the BS law, due to the global induction 
effects of the vortex. This unexpected result has motivated further 
work and some current research which is still in progress.

\begin{figure}[t!]
\centering
\includegraphics[width=\textwidth]{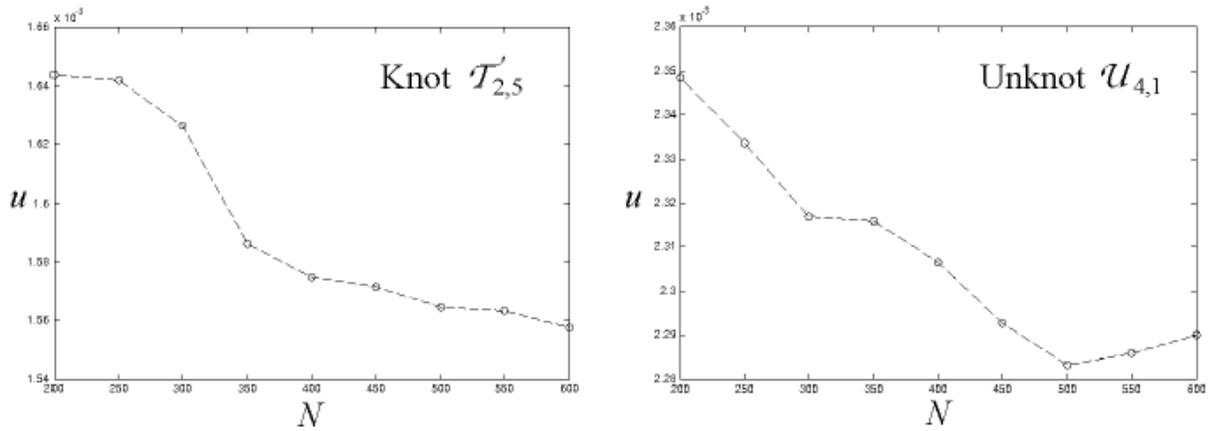}
\caption{Examples of convergence test given by plots of the 
translational velocity $u$ against the number of discretization 
points $N$ on the vortex axis. Similar results have been found for the 
other knots/unknots tested. Broken lines are for visualization purposes 
only.}
\label{convergence}
\end{figure}

\section{Numerical method}
\label{sec:NumericalMethod}
Dynamical quantities of vortex knots and unknots are 
evaluated by direct numerical integration of the BS law (\ref{bs}) 
and direct numerical calculation of the other properties. 
The numerical code is described in detail elsewhere \cite{ref:Sultan} 
and it has been used also to study interaction and reconnection of vortex 
bundles \cite{ref:Aea08}. The vortex axis is discretized into $N$ segments 
and the Biot-Savart integral is de-singularized by application of a 
standard cut-off technique \cite{ref:Sch88,ref:Aa94,ref:Sultan}. The time 
evolution is realized by using a $4^{th}$ order Runge-Kutta algorithm.
Convergence has been tested in space and time as usual by modifying 
the number of discretization points and the size of the time step.

\begin{figure}[t!]
\centering
\includegraphics[width=\textwidth]{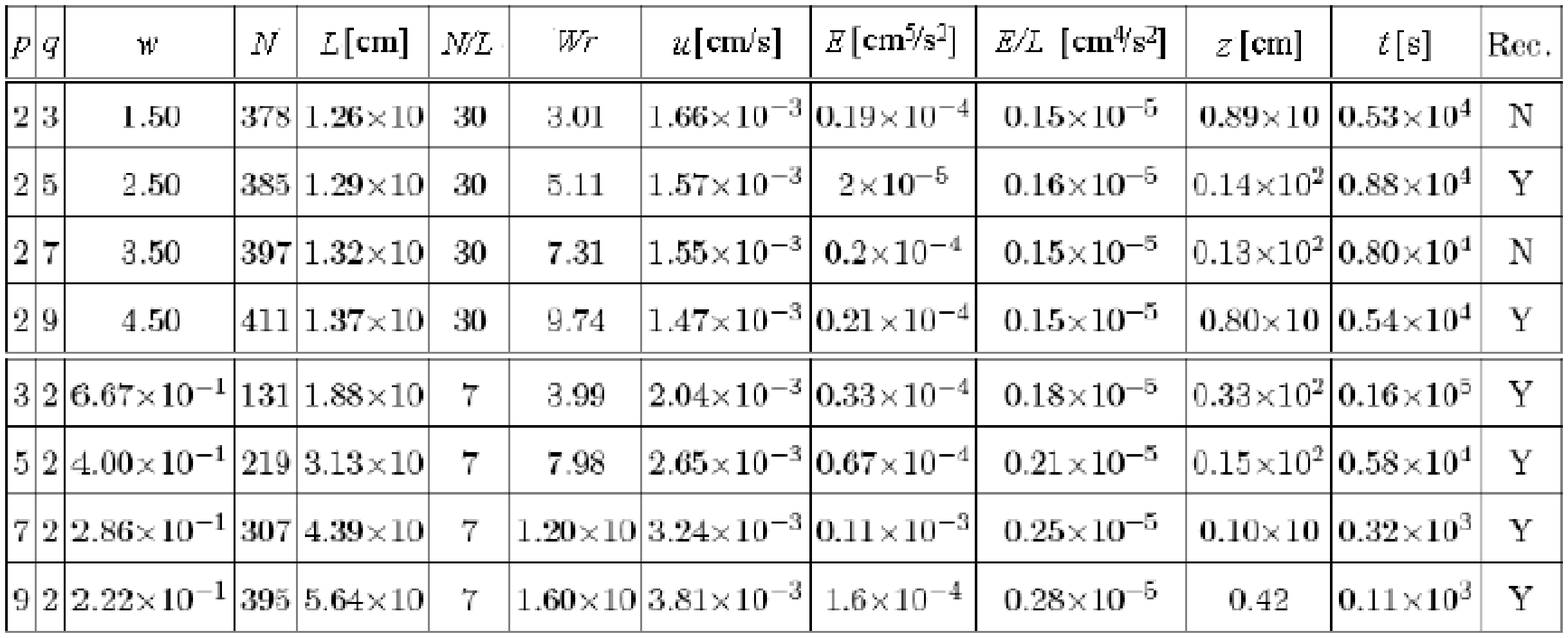}
\caption{Vortex knots: numerical values of the calculated quantities.
Entries in the last column report whether (Y) or not (N) a reconnection 
event has taken place on or before time $t$.}
\label{tableone}
\end{figure}

In all calculations, the initial condition is given by (\ref{sols}).
We set $r_0=1~\rm cm$, $\epsilon=0.1$, $\Gamma=10^{-3}~\rm cm^2/s$, 
(the value expected for superfluid $^4$He), $\delta=2\times 10^{8}/e^{1/2} 
\gg 1$ (typical of a superfluid helium vortex core radius $a=10^{-8} \rm cm$).
To analyse unknots we replace in eqs. (\ref{sols}) $(1-1/w^2)$ with 
$|1-1/w^2|$. It is useful to compare these results with the dynamics of 
a vortex ring $\cU_{0}$ of same size and vorticity. Thus we take a reference 
vortex ring with radius $r_0=1~\rm cm$.  
Typical time-step value in our calculations is $10^{-2}~\rm s$. 
Convergence in time has been tested by ranging the time-step 
from $10^{-3}~\rm s$ to $5 \times 10^{-2}~\rm s$.
Examples of convergence in space are shown in Figure~\ref{convergence}
for torus knot $\mathcal{T}_{2,5}$ and 
unknot $\mathcal{U}_{4,1}$. The calculations are performed using 
constant mesh density $N/L$ chosen following convergence tests similar 
to those shown in Figure~\ref{convergence}. We set $N/L=50$ for poloidal unknots $\mathcal{U}_{1,m}$ ($m=2,3,\ldots,7$), $N/L=20$ for toroidal unknots 
$\mathcal{U}_{m,1}$ ($m=2,3,\ldots,7$), $N/L=30$ for  knots 
$\mathcal{T}_{2,q}$ ($q=3,5,7,9$) and $N/L=7$ for  knots 
$\mathcal{T}_{p,2}$ ($p=3,5,7,9$). Typical errors in computing the 
velocity and the energy are approximately $10^{-5}~\rm cm/s$ and 
$10^{-7}~\rm cm^5/s^2$, respectively. For the reference vortex ring we 
set the winding number $w=1$ and $N=313$; then $L=6.26~\rm cm$, thus 
$N/L=50$, translational velocity $u_0=1.38 \times 10^{-3}~\rm cm/s$, 
kinetic energy $E_0=8.77 \times 10^{-6}~\rm~cm^5/s^2$ and obviously 
$H=0~\rm cm^4/s^2$.

\begin{figure}[t!]
\centering
\includegraphics[width=\textwidth]{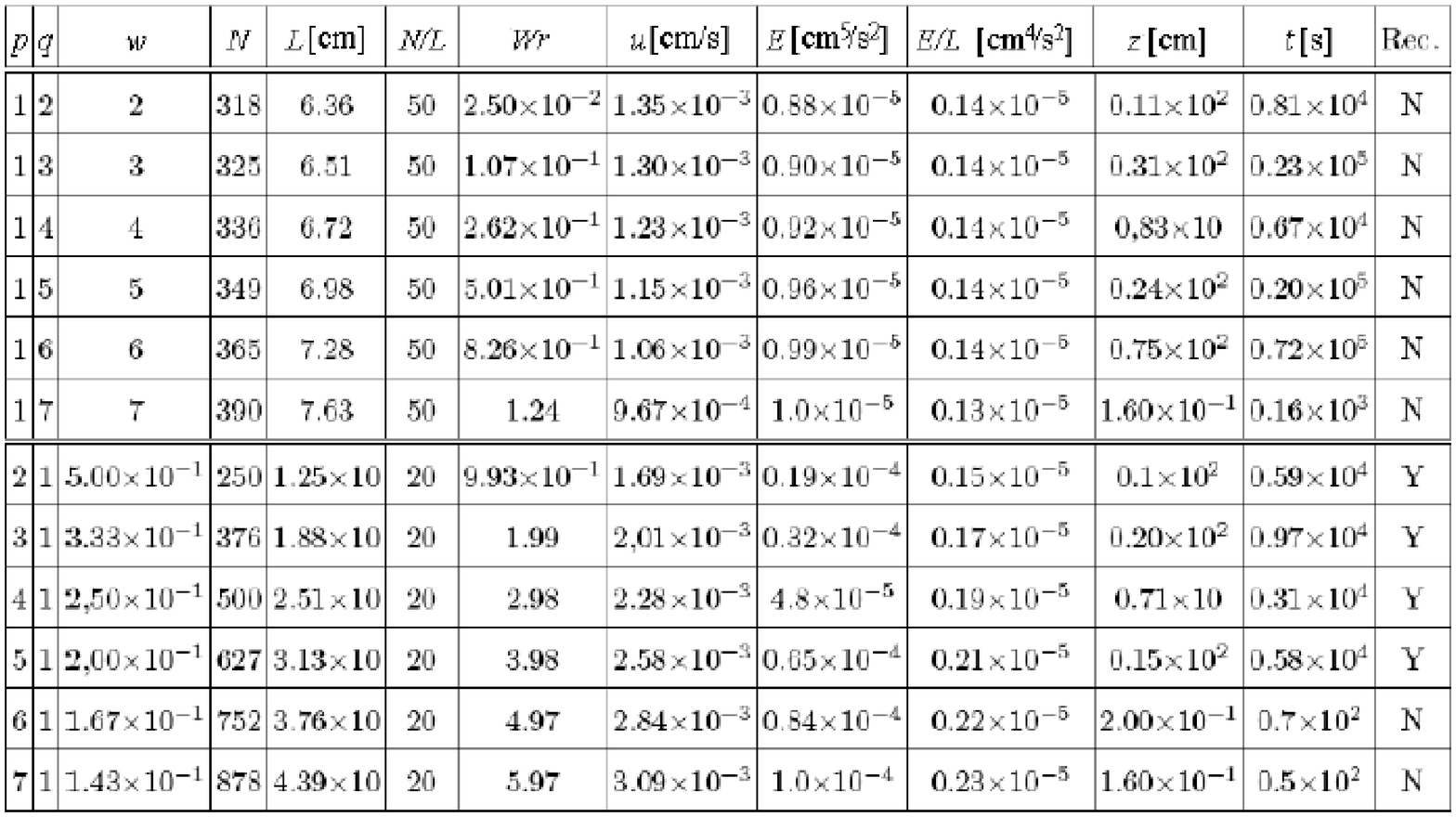}
\caption{Vortex unknots: numerical values of the calculated quantities.
Entries in the last column report whether (Y) or not (N) a reconnection 
event has taken place on or before time $t$.}
\label{tabletwo}
\end{figure}

\section{Results: velocity, energy and helicity}
\label{sec:NumericalResults}
Numerical values of calculated quantities are reported in the tables 
of Figure~\ref{tableone} and \ref{tabletwo}. Let us consider first 
purely geometric information, that will be useful to understand the 
following results on velocity, energy and helicity.

\subsection{Total length and writhing number}
\begin{figure}[t!]
\centering
\includegraphics[width=\textwidth]{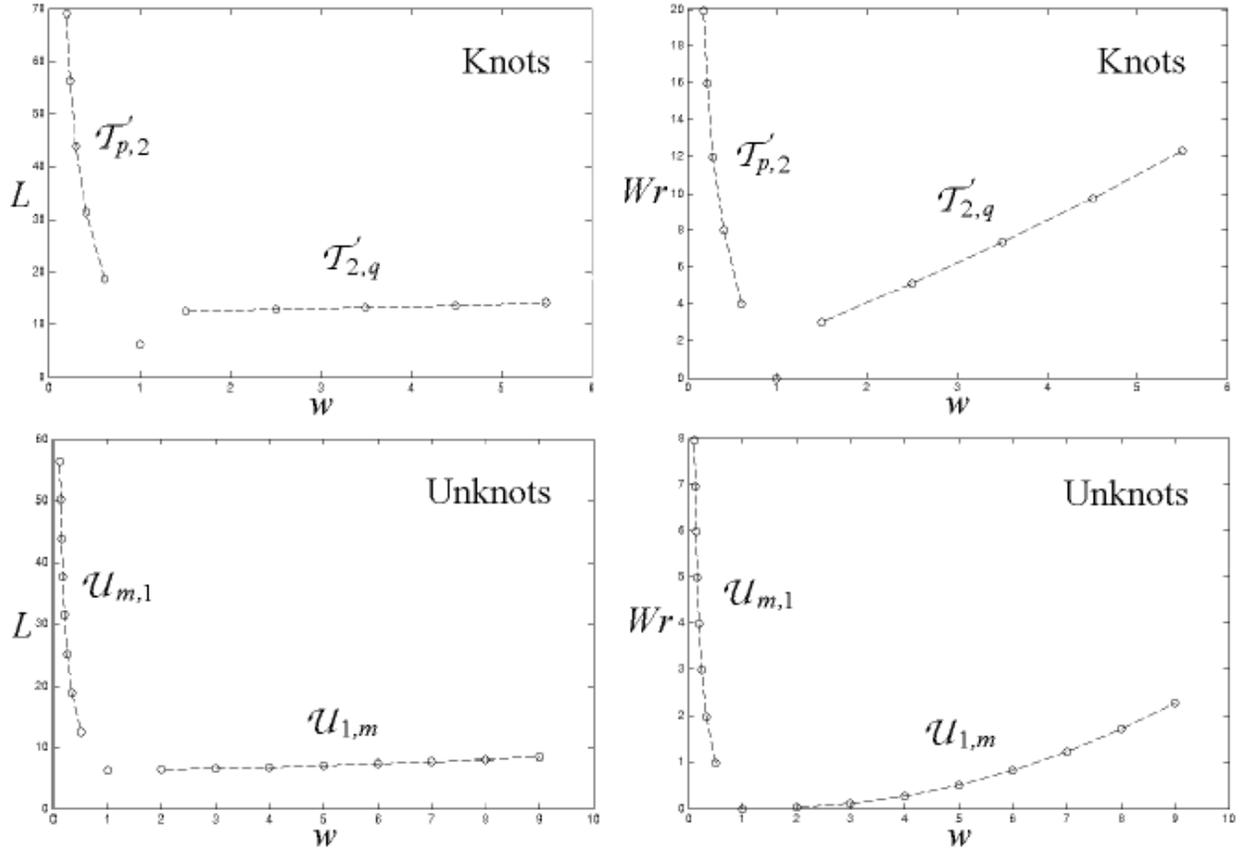}
\caption{Total length $L$ and writhing number $Wr$ plotted against
the winding number $w$ of the torus knots and unknots calculated
($p=3,5,7,9$; $q=3,5,7,9$; $m=2,3,\ldots,7$).
The isolated circle denotes the reference vortex ring value. 
Broken lines are for visualization purposes only.}
\label{geometry}
\end{figure}

Diagrams of total length and writhing of knots and unknots are shown in 
Figure~\ref{geometry}. From diagrams of total length (left column), we 
see that these plots reflect the elementary fact that longitudinal wraps
contribute to total length more than meridian wraps. The marked difference 
in the slope of the two plots in each diagram is just due to the dominant 
contribution to $L$ by the longitudinal wraps compared to the modest one 
of the meridian wraps. This behavior will be reflected in 
the relative kinetic energy of the system. 

Similarly for the amount of coiling and distortion of the filament axis
in relation to the dynamics and helicity of the vortex. As we can see 
from the diagrams on the right of Figure~\ref{geometry}, meridian wraps
contribute modestly (if not at all appreciably to our degree of accuracy) 
to the total writhing of the filament, the dominant contribution coming 
from the longitudinal wraps present. Moreover, since the self-linking of 
the vortex filament is a topologically conserved quantity \cite{ref:RM92} 
given by $Lk=Wr+Tw$, where $Tw$ denotes total twist of the filament, 
then information on writhing provides direct information on total twist, 
by taking $Tw=Lk-Wr$. Furthermore, being the number of meridian wraps
directly proportional to total twist, we have an immediate estimate of 
the relative contributions to helicity (see the discussion in the 
sub-section below).

\subsection{Translation velocity}
\begin{figure}[t!]
\centering
\includegraphics[width=0.6\textwidth]{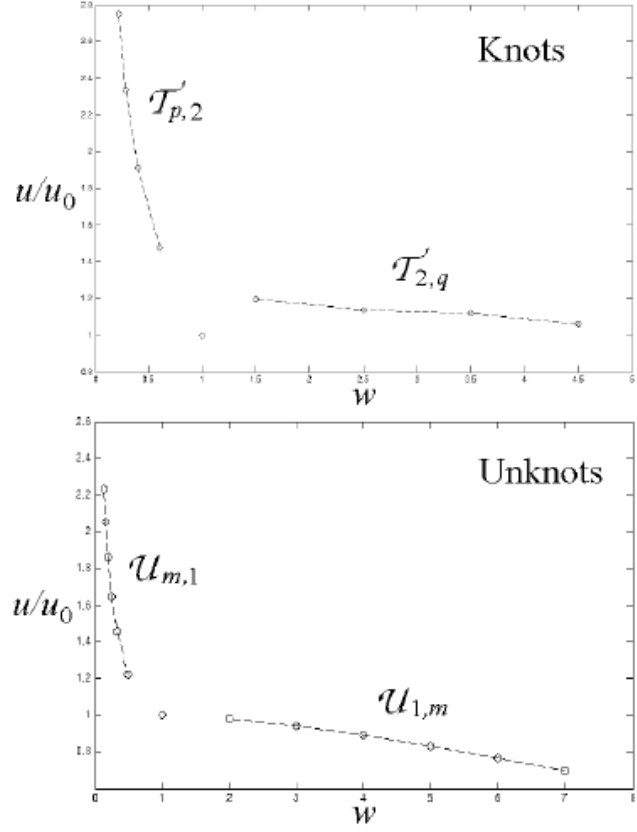}
\caption{Translational velocity $u$ normalized with respect to the 
velocity $u_0$ of a reference vortex ring plotted against
the winding number $w$ of the torus knots and unknots calculated
($p=3,5,7,9$; $q=3,5,7,9$; $m=2,3,\ldots,7$). 
The isolated circle denotes the reference vortex ring value. 
Broken lines are for visualization purposes only.}
\label{velocity}
\end{figure}

The translational velocity $u=\dot z$ along the central axis of knots 
and unknots is calculated by using the Biot-Savart law.  
Absolute values are reported in the tables of Figure~\ref{tableone} and
\ref{tabletwo}. The diagrams of Figure~\ref{velocity} show the normalized 
velocity $u/u_0$ of knots and unknots plotted against the winding number.
The velocity is greatly influenced by the relative number of longitudinal 
wraps, which contribute greatly to the total curvature of the vortex. In 
general the velocity decreases with increasing winding number. Fastest 
vortex systems are thus torus knots with highest number of longitudinal 
wraps. In the case of unknots we can see that meridian wraps actually 
slow down poloidal coils $\cU_{1,m}$, making them traveling slower than 
the corresponding vortex ring. At very high winding number, torus 
knots and poloidal coils seem to reverse their velocity, thus 
traveling backward in space. This curious phenomenon has been observed 
independently \cite{ref:KiMa02,ref:Ba06}, and it can be justified on 
theoretical grounds by information based on structural complexity analysis 
\cite{ref:Ri08,ref:Ri09}.

An estimate of the relationship between normalized velocity $u/u_0$ and
winding number $w$ of torus knots is obtained by a linear regression, 
given by:
\begin{equation}
u/u_0 =
\left\{ 
\begin{array}{l}
4.41 - 9.13 w +7.09 w^2\ , \qquad (w<1) \ , \\[1mm]
1.25 - 0.04 w\ , \qquad (w>1) \ ,
\end{array} 
\right.
\label{regression1}
\end{equation} 
with standard deviation of 0.031.

In the limit $w\to0$, the knot covers the toroidal surface completely
with an infinite number of longitudinal wraps, and vorticity becomes 
a sheet of toroidal vorticity, with the induced velocity, purely poloidal 
in the interior and exterior region of the torus, that jumps across the 
sheet in opposite directions. The regression (\ref{regression1}) suggests 
a theoretical limit value $u/u_0 =4.41$ that seems to be independent of 
the aspect ratio of the torus. In the other limit, $w\to\infty$, the knot 
covers the torus with an infinite number of meridian wraps, and in this 
case vorticity induces a toroidal jet in the interior region. 

\begin{figure}[t!]
\centering
\includegraphics[width=\textwidth]{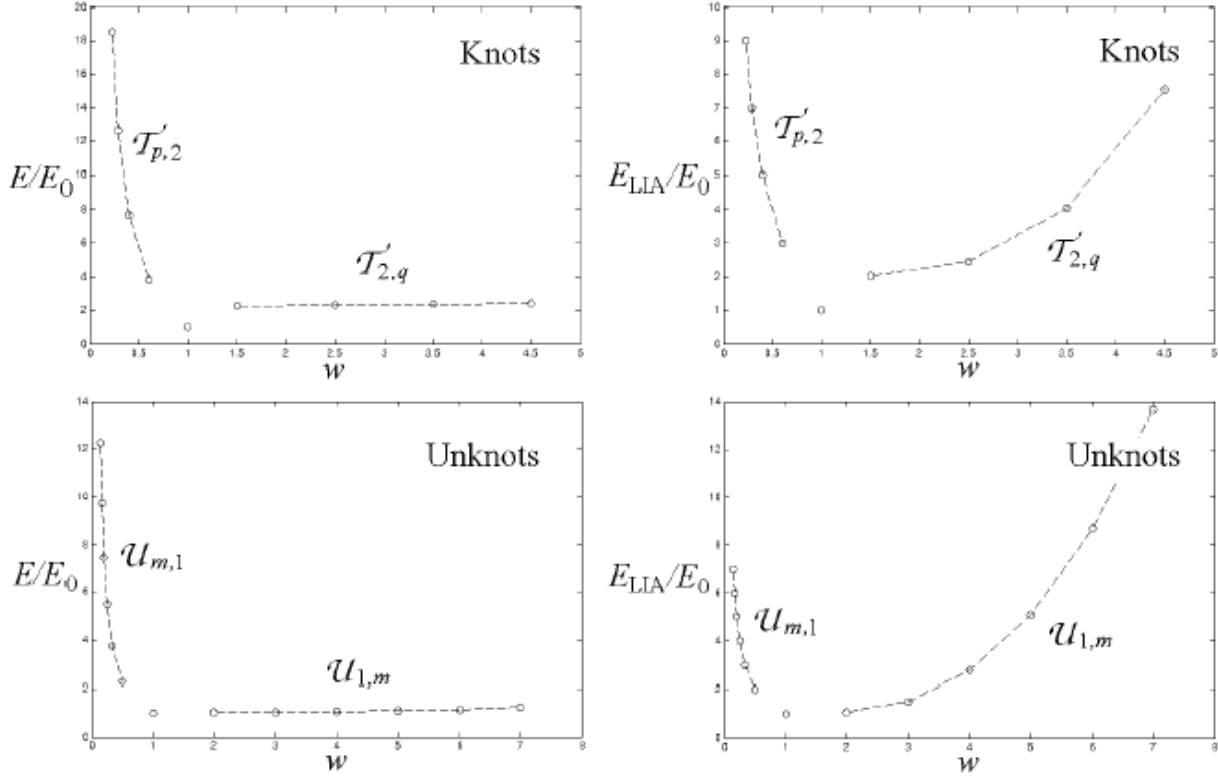}
\caption{Left column: kinetic energy $E$ calculated by using the Biot-Savart 
law and normalized with respect to the energy $E_0$ of the reference vortex 
ring, plotted against the winding number $w$ of the knots and unknots 
calculated ($p=3,5,7,9$; $q=3,5,7,9$; $m=2,3,\ldots,7$). Right column:
kinetic energy $E_{\rm LIA}$ calculated by using the LIA law and normalized 
with respect to the energy $E_0$ plotted against $w$. The isolated circle 
denotes the reference vortex ring value. Broken lines are for visualization 
purposes only.}
\label{energy}
\end{figure}

\subsection{Kinetic energy}
\label{sec:KineticEnergyOfTorusKnotsAndUnknots}
The diagrams of the normalized kinetic energy per unit density $E/E_0$
(Figure~\ref{energy}, left column) of the knots and unknots calculated 
by using the Biot-Savart law, and the normalized energy calculated by 
the LIA law $E_{\rm LIA}/E_0$ (Figure~\ref{energy}, right column), plotted 
against the winding number $w$, show trends dictated by the behavior of 
total length $L=L(w)$ (cf. the diagrams of Figure~\ref{geometry}). Here
the calculation of the volume integral (\ref{kinetic}) is replaced by the 
more economical line integral \cite{ref:BRS01}
\begin{equation}
E=\frac{\Gamma}{2}\oint_{\cC}\bu\cdot\bX\times\bt\, \rd s\ .
\label{kineticline}
\end{equation}
An estimate of the relationship between the normalized energy $E/E_0$ 
of torus knots and their winding number $w$ is given by a non-linear 
regression based on least squares:
\begin{equation}
E/E_0 = 2.35 + 67.03 e^{- 6.44 w}\ , \qquad (w<1) \ ,
\end{equation}
with root mean square deviation of 0.218.

Different trends and values are obtained by calculating the normalized 
energy by using the LIA law; in this case, by using (\ref{lia}), we have
\begin{equation}
E_{\rm LIA}=\frac{1}{2}\int_{V}\|\bu_{\rm LIA}\|^2\,\rd^{3}\bx
=\left(\frac{\Gamma\ln\delta}{4\pi}\right)^2\oint_{\cC} c^{2}\,\rd s\ ,
    \label{kineticlia}    
\end{equation}
that is one of the conserved quantities associated with the LIA law 
\cite{ref:Ri93}. Direct comparison between the diagrams of the two 
energies reveals two distinct trends: for $w<1$ the LIA law 
under-estimates the actual energy of the vortex (knotted or unknotted),
whereas for $w>1$ the LIA provides much higher energy values. The 
importance of non-local effects is evident: the LIA energy of $\cT_{9,2}$ 
and $\cU_{7,1}$, for example, is about 40\% less than the corresponding 
BS energy, whereas $\cT_{2,9}$ and $\cU_{1,7}$ under LIA have 3 and 14 
times more energy than their corresponding BS counterparts. These 
differences are essentially due to the contributions from the induction
effects of nearby strands, captured by the BS law, but completely 
neglected under LIA. 

Another quantity of interest is the kinetic energy per unit length $E/L$.
From the tables of Figures \ref{tableone} and \ref{tabletwo} we see that 
for $w<1$ $E/L$ can vary by up to 100\%. This result has interesting 
implications for the interpretation of experiments on superfluid turbulence, 
particularly at low temperatures, a regime that is dominated by Kelvin waves 
\cite{ref:KVSB01}. What is measured in the experiments is the total vortex 
length per unit volume $\Lambda$, and the turbulence energy per unit volume 
is deduced by multiplying the observed vortex length per volume times the 
energy per unit length; by integrating the velocity field of a straight 
vortex filament, the energy per unit length is estimated as 
$E/L \approx \Gamma^2/(4 \pi^2) \ln{(b/a)}$, where $b \approx \Lambda^{-1/2}$ 
is the typical distance between vortices. Our results show that for highly 
bent vortex filaments (as in the case of superfluid turbulence, where 
vorticity is even fractal \cite{ref:fractal}), $E/L$ is certainly not 
constant.

\begin{figure}[t!]
\centering
\includegraphics[width=0.6\textwidth]{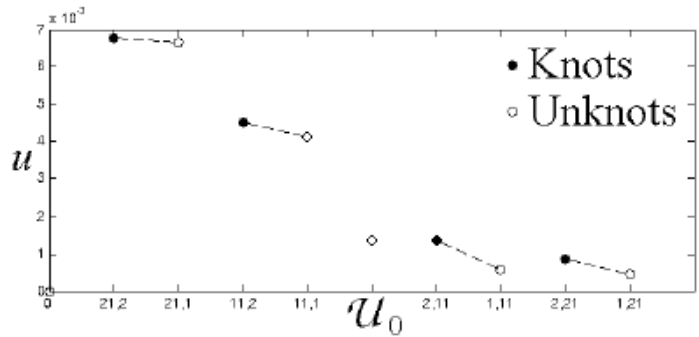}
\caption{Translational velocity $u$ of some knots and related unknots 
reported here for comparison. In general knots travel faster than their 
corresponding unknots. The isolated circle denotes the reference vortex 
ring value. Broken lines are for visualization purposes only.}
\label{topology}
\end{figure}

\subsection{Helicity}
\label{sec:helicity}
It is interesting to investigate the effects of topology on the 
velocity and kinetic helicity, by exploring the interplay of geometric
and topological aspects on the dynamics. For this let us consider 
Figure~\ref{topology}, where the velocities of a torus knot and that of 
the corresponding unknot with same number of longitudinal or meridian 
wraps are shown for comparison. As we see in general vortex knots travel 
faster than their corresponding unknots, and the higher the number of 
longitudinal wraps the faster is the motion, whereas the higher the number 
of meridian wraps the slower is the propagation speed. 

As far as helicity is concerned, we have $H=\Gamma Lk$ \cite{ref:RM92} 
and for torus knots we can set $Lk=pq$, that is invariant for each 
knot type ($Tw$ being the total twist) and it increases with knot complexity. 
Since $Lk=Wr+Tw$ from the $Wr$-values of the table of Figure~\ref{tableone} 
we can compute $Tw$ for each knot. In general twist helicity is larger for 
knots with higher number of meridian wraps.
For the unknots we can set $Lk=Wr+Tw=0$  (for the reference vortex ring we have 
zero writhe and zero twist); thus from the values of the table of 
Figure~\ref{tabletwo}, we have $Tw=-Wr$ and therefore we can estimate writhe 
and twist helicity contributions.  

\begin{figure}[t!]
\centering
\includegraphics[width=\textwidth]{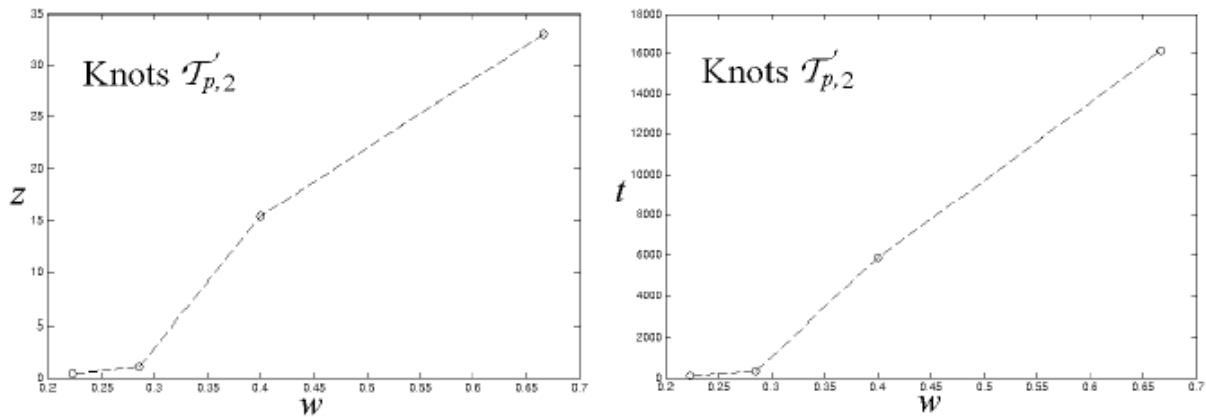}
\caption{Plots of space travelled ($z$) and time elapsed ($t$) before a 
reconnection event takes place for torus knots $\cT_{p,2}$ ($p=3,5,7,9$) 
plotted against the winding number $w$. Calculations are based on the 
Biot-Savart law. Broken lines are for visualization purposes only.}
\label{reconnection}
\end{figure}

\subsection{Structural stability considerations}
\label{sec:structuralstability}
Aspects of structural stability of vortex knots and unknots based on 
permanence of knot signature and occurrence of a reconnection event
are explored. In previous work we noticed the stabilizing effect 
of the BS law on LIA-unstable knots \cite{ref:RSB99}. Here we extend 
this comparison to the knots and unknots considered so far. 
In general a vortex structure is said to be stable if it evolves under 
signature-preserving flows that conserve topology, geometric signature 
and vortex coherency. Let us consider results reported in the tables of 
Figure~\ref{tableone} and \ref{tabletwo} and shown in 
Figure~\ref{reconnection}. 

Since our results concern both Euler's and superfluid dynamics, it is 
important to remark that superfluid vortices can reconnect with each other 
in absence of dissipation \cite{ref:KoLe93,ref:BPSL09}, whereas in the Euler 
context vortex topology is conserved; this important difference has been 
recently reviewed by Barenghi \cite{ref:B08}. Here we adopt the following 
criterium: when, during the evolution, a vortex reconnection takes place, 
then we stop the calculation, and deem the structure to be unstable.
The last column of the table of Figure~\ref{tableone} and \ref{tabletwo}
refers to the occurrence of a reconnection at instant $t$. 
In the absence of reconnection, we report the distance $z$ travelled by 
the vortex during the computational time $t$. If $z$ is much larger than 
the typical vortex size, the knot/unknot tested is said to be structurally 
stable.

From the last two columns of Figure~\ref{tableone} we see that for $w<1$
the space travelled tends to decrease with increasing knots complexity; 
for example, $t(\cT_{3,2}) \gg t(\cT_{9,2})$. The effect is also shown 
in Figure~\ref{reconnection}. Note that these knots would be LIA-unstable. 
Thus, the stabilizing effect due to the BS law is confirmed: these knots 
can indeed travel a distance which is larger than their size before 
unfolding and reconnecting. 

\section{Conclusions}
In this paper we examined the effect of several geometric and topological
aspects on the dynamics and energetics of vortex torus knots and unknots 
(i.e. toroidal and poloidal coils). This study is carried out by numerically 
integrating the Biot-Savart law, and by comparing results for several knots 
and unknots at different winding numbers $w$. Generic behaviors are found for 
the class of knots/unknots tested, and main results are presented by normalizing
velocity and energy by the corresponding value of a standard vortex 
ring ($\cU_0$) of same size and circulation. 

In general, for $w<1$ (where the number of longitudinal wraps is larger 
than that of meridian wraps) the more complex the vortex structure is, the 
faster it moves, and both torus knots and toroidal coils move faster than 
$\cU_0$. For $w>1$ (where the relative number of meridian wraps dominates) 
all vortex structures move essentially as fast as $\cU_0$, almost independently 
from their total twist. Therefore, for all the structures tested total twist 
provides only a second-order effect on the dynamics.

We have also found that for $w<1$ vortex structures carry more kinetic energy 
than $\cU_0$, whereas for $w>1$ knots and poloidal coils have almost the same 
energy as $\cU_0$. The LIA law (an approximation often used to replace the 
Biot-Savart law, more computationally demanding) under-estimates the energy of 
knots with $w<1$ and over-estimate the energy for $w>1$.

Kinetic helicity, that is naturally decomposed in writhe and twist 
contributions, is evidently determined by the relative number of longitudinal 
and meridian wraps present, the latter contributing to twist helicity rather 
modestly (relatively). 

Finally, we can extend previous results and confirm that for $w<1$ the 
Biot-Savart law has a stabilizing effect on knots that are LIA-unstable:
all vortex structures tested have been found to be structurally stable 
regardless of the value of $w$, being able to travel in the fluid for 
several diameters before eventually unfolding and reconnecting. 


\end{document}